\documentclass[11pt]{article}
\usepackage{amsmath}
\usepackage{amssymb}
\usepackage{amsfonts}
\usepackage{amscd}
\usepackage{accents}
\usepackage{pst-all}
\usepackage{epsfig}
\usepackage{graphicx}
\usepackage[tight]{subfigure}
\date{\today}
\newcommand{\bmat}{\left(\begin{array}}
\newcommand{\emat}{\end{array}\right)}
\newcommand{\be}{\begin{equation}}
\newcommand{\ee}{\end{equation}}
\newcommand{\ba}{\begin{eqnarray}}
\newcommand{\ea}{\end{eqnarray}}

\def\lsim{\raise0.3ex\hbox{$\;<$\kern-0.75em\raise-1.1ex\hbox{$\sim\;$}}}
\def\gsim{\raise0.3ex\hbox{$\;>$\kern-0.75em\raise-1.1ex\hbox{$\sim\;$}}}

\def\be{\beta}

\begin{document}

\vspace*{-.6in} \thispagestyle{empty}
\begin{flushright}
LPT-Orsay-14-14
\end{flushright}
\baselineskip = 20pt

\vspace{.5in} {\Large
\begin{center}
{\bf  Axial Dark Matter: the case for an invisible Z$^\prime$  
  }
\end{center}}

\vspace{.5in}

\begin{center}
{\bf  Oleg Lebedev$~^1$ and  Yann Mambrini$~^2$   }  \\

\vspace{.5in}

$^1$\emph{Department of Physics and Helsinki Institute of Physics,
Gustaf H\"allstr\"omin katu 2, FIN-00014 University of Helsinki, Finland
 }\\
$^2$\emph{Laboratoire de Physique Th\' eorique Universit\' e Paris-Sud, F-91405 Orsay, France 
}
\end{center}

\vspace{.5in}

\begin{abstract}
\noindent
We consider the possibility that fermionic dark matter (DM)  interacts with the Standard Model fermions through an axial Z$^\prime$ boson. 
 As long as  Z$^\prime$ decays predominantly into dark matter, the relevant LHC bounds are rather loose. 
 Direct dark matter detection does not significantly constrain   this scenario either, since dark matter scattering on nuclei is spin--dependent. As a result, for a range of the Z$^\prime$ mass and couplings,
the DM annihilation cross section is large enough to be consistent with thermal history of the Universe. 
In this framework, the thermal WIMP paradigm, which currently finds itself under pressure,  is perfectly viable. 
 \end{abstract}

 \newpage
 
\section{Introduction}

Models with an extra U(1) are among the simplest and most natural 
extensions of the Standard Model (SM).  They enjoy both the top--down
and bottom--up motivation. In particular, additional U(1)'s appear in many 
string constructions. From the low energy perspective, the coupling 
between an SM fermions $f$ and a massive gauge boson Z$^\prime$ \cite{Langacker:2008yv}
\begin{equation}
{\cal{L}}_{\rm int}=  g_f ~ Z^\prime_\mu ~  \bar f \gamma^\mu (a+b \gamma^5) f ~,
\end{equation}
where $g_f,a,b$ are some constants,
represents one of the dimension-4 ``portals'' (see e.g.~\cite{Domingo:2013tna}) connecting the observable world
to the SM--singlet sector. 
This is particularly important in the context of dark matter models
\cite{Cheung:2007ut}.
If dark matter is charged under the extra U(1),
the above coupling  provides a DM annihilation channel into visible particles. As long as 
the  Z$^\prime$ has a TeV scale mass and the couplings are not too small,
this framework fits the WIMP--miracle paradigm \cite{Bertone:2004pz}.
Recent LHC  \cite{Chatrchyan:2012oaa,Aad:2012hf} and direct DM detection constraints \cite{Akerib:2013tjd}, however, put significant
 pressure on this idea since no traces of a Z$^\prime$ were found in 
 either direct collider searches or DM scattering on nuclei.

In 
this letter, we argue that these negative results may be due to the
axial nature of  the Z$^\prime$ and its stronger coupling to dark matter compared to $g_f$ above. In this case, which we call ``axial dark matter''
(AxDM), DM scattering on nuclei is spin--dependent and weakly constrained. The LHC has limited sensitivity to 
such a Z$^\prime$ due to the fact that it decays predominantly into dark matter, as in \cite{Arcadi:2013qia}.\footnote{We allow a Z$^\prime$ to couple
  universally to SM fermions, which distinguishes the model from the leptophobic scenarios (see e.g.~\cite{Alves:2013tqa}).}
We thus find that all of the constraints can be satisfied, which adds some
credibility to the WIMP paradigm.

\section{Axial Z$^\prime$  }

In what follows, we consider the possibility that Z$^\prime$ is purely axial, with the couplings\footnote{An analysis of the axial DM coupling to the 
usual Z--boson has recently appeared in \cite{deSimone:2014pda}.}
\begin{equation}
{\cal{L}}_{\rm int}^{\rm eff}= \sum_f  g_f ~ Z^\prime_\mu ~  \bar f \gamma^\mu \gamma^5 f +
g_\chi ~ Z^\prime_\mu  ~ \bar \chi \gamma^\mu \gamma^5 \chi~.
\end{equation}
Here $f$ represents the Standard Model (SM) fermions, $\chi$ is a 
Dirac fermion constituting dark matter and $ g_f, g_\chi$ are the 
corresponding  Z$^\prime$ couplings. 
This Lagrangian represents an effective low energy interaction
after heavy particles have been integrated out and the vector boson kinetic terms have been diagonalized. Clearly, the microscopic theory can be made anomaly--free by assigning appropriate charges to fermions (we do not exclude the possibility of having further heavy fermions coupled to Z$^\prime$). 

One may ask how natural it is to have a pure axial--vector interaction.
In our opinion, this choice is quite natural given the fact that the photon interaction is purely vector and the axial case is just the other extreme. Also,
our considerations  hold in the presence of a small vector component of  Z$^\prime$, which may be generated through kinetic mixing \cite{Holdom:1985ag}.

To make our model as simple as possible, we will focus on the case of a universal coupling of    Z$^\prime$ to the SM fermions, $g_f$. (This assumption can of course be easily relaxed by inserting the fermion--dependent charges.) We then find that cosmological and accelerator constraints require 
\begin{equation}
g_f \ll g_\chi ~, \label{hierarchy}
\end{equation}
by a factor of ${\cal O}(10)$ to ${\cal O}(10^3)$. 
One would be hesitant to attribute such a hierarchy to the difference in the observable and hidden charges. On the other hand, factors of this type can arise in the system of two U(1)'s mixing with each other.
Consider the general Lagrangian describing two massive abelian gauge bosons,
 \begin{equation}
 {\cal{L}}_{AB}= -{1\over 4} (F_A^{\mu\nu})^2 - {a\over 2} F_A^{\mu\nu}
 F_{B\mu\nu} -{1\over 4} (F_B^{\mu\nu})^2
 + {1\over 2} M_1^2 A_\mu^2 + \delta M^2 A_\mu B^\mu + {1\over 2} M_2^2 B_\mu^2 ~,
 \end{equation}
where $A$ couples only to the dark sector with coupling $g_A$, while
$B$  couples only to the visible sector with coupling $g_B$.
The lighter mass eigenstate would be a mixture of $A$ and $B$, which
couples to both sectors. The hierarchy (\ref{hierarchy}) can then be
recovered in various limits. For example, it can result from 
$M_1^2 \ll M_2^2 $. For order one kinetic mixing, $a\sim 1$, the  Z$^\prime$
is composed mostly of $A$ and 
\begin{equation}
g_\chi =  {\cal{O}}(g_A ) ~~, ~~ 
 g_f =  {\cal{O}}\left( {M_1^2 \over M_2^2} ~ g_B \right) ~.
\end{equation}
Another possibility is to attribute  (\ref{hierarchy}) to the 
hierarchy in the couplings, $g_A  \gg g_B $.
For a small kinetic mixing $a \sim 0$ and large mass mixing 
$ M_1^2 \sim  M_2^2 \sim  \delta M^2$, the Z$^\prime$ is a mixture of 
$A$ and $B$ with 
\begin{equation}
g_\chi =  {\cal{O}}(g_A ) ~~, ~~ 
 g_f = {\cal{O}}\left( g_B \right) ~.
\end{equation} 
Note that for $ M_1^2 \approx M_2^2 \approx  \delta M^2$ ,
the mixing is nearly maximal and 
 the second mass eigenstate becomes heavy.
In what follows, we will be agnostic as to the origin of the hierarchy
(\ref{hierarchy}) and will treat the two couplings as free parameters.

\section{Dark matter and Z$^\prime$ phenomenology}

In this section, we provide a list of cosmological and accelerator constraints on the model. These set bounds on the two couplings $g_f , g_\chi$ and the Z$^\prime$ mass $m_{Z^\prime}$. In order to understand their
qualitative behaviour and compatibility, we provide simple analytic approximations for  the observables.  

\noindent {\bf Planck/WMAP and DM annihilation.}
Suppose that DM is produced thermally, as in the traditional WIMP 
scenario. 
The main dark matter annihilation mechanism is the $s$-channel annihilation into SM fermion pairs.
Although we will use the exact tree--level result in our numerical analysis, it is instructive to consider the heavy Z$^\prime$,
$m_{Z^\prime}^2 \gg m_\chi^2$, and  zero DM--velocity limit.\footnote{Numerically, the velocity--independent terms dominate at relatively low $m_{Z^\prime}$, while for a heavier Z$^\prime$ velocity--dependent contributions  are equally important. We choose the limit $v\rightarrow 0$ 
for transparency of our discussion, while using the full result in our numerical analysis. }
In this case, the cross section takes on a 
particularly simple form,
\begin{equation}
\langle \sigma v \rangle = { g_f^2 g_\chi^2 \over 2 \pi } ~c_f ~
\sqrt{ 1- { m_f^2 \over m_\chi^2}} ~ {m_f^2 \over m_{Z^\prime}^4 } ~,
\end{equation}
where $c_f$ is the number of colors for quarks and 1 for leptons. We see  that, for light final state fermions, the cross section is suppressed.
The origin of the $ m_f^2 / m_{Z^\prime}^2$ factor can be understood
from (conserved)  C-parity considerations. The C-parity of the initial state must be +1 to match that of Z$^\prime$. Since for a fermion--antifermion pair it is given by $(-1)^{l+s}$ with $l$ and  $s$ being the angular momentum
and spin quantum numbers, the
$s$-wave initial state ($v \rightarrow 0$)  must then have  $s=0$.
On the other hand, the helicities of the  relativistic final state  fermions
add up to 1. Hence, a spin flip is required leading to the 
 $ m_f / m_{Z^\prime}$ dependence.  
Note however that, for heavy fermions like the top quark, this 
factor does not lead to significant suppression of the amplitude.

Suppose that DM is sufficiently heavy such that its pair annihilation into top quarks is allowed. Then for $\sqrt{ 1- { m_t^2 / m_\chi^2}}  \sim 1$,
the canonical WIMP annihilation cross section $\sigma v = 
3 \times 10^{-26}$ cm$^3$s$^{-1}$ translates into
\begin{equation}
 {m_{Z'} \over \sqrt{ g_f g_\chi} } \sim  1500 ~{\rm GeV} ~.
 \label{annih}
 \end{equation}
 One should keep in mind that this figure indicates the ballpark of the result   and the 
 velocity- as well as $m_\chi$--dependent contributions affect $\langle \sigma v \rangle$, while the definitive
 answer is given by our numerical analysis. 
 \\ \ 
\noindent {\bf Direct DM detection.}
Tree level Z$^\prime$ exchange leads to  spin--dependent DM scattering
on nuclei, which is constrained by a number of experiments.
For an (approximately) universal Z$^\prime$ coupling to quarks,
\begin{equation}
\sigma^{\rm SD}= { 12 g_f^2 g_\chi^2  m_N^2 \over  \pi
m_{Z^\prime}^4  }  ~
(\Delta_u + \Delta_d + \Delta_s)^2  ~,
\end{equation}
where $m_N$ is the nucleon mass and $\Delta_i$ are the quark 
contributions to the proton spin: $\Delta_u=0.84$, 
 $\Delta_d=-0.43$, $\Delta_s=-0.09$. Taking $10^{-39}$ cm$^2$ as the benchmark bound on $\sigma^{\rm SD}$ for $m_\chi \sim 100$ GeV 
 \cite{Aprile:2013doa}, one finds 
 \begin{equation}
 {m_{Z'} \over \sqrt{ g_f g_\chi} } > 600 ~{\rm GeV} ~.
 \end{equation}
This bound is satisfied automatically for thermally produced dark matter
(see  Eq.~(\ref{annih})).

Spin--independent DM scattering  is generated at one loop with the corresponding
amplitude being suppressed both by a loop factor and the quark masses
required by a helicity flip. The resulting bound is weak  \cite{Freytsis:2010ne}.  

We note that similar conclusions apply
to the DM--nucleon interaction mediated by a pseudoscalar 
as recently studied in \cite{Boehm:2014hva}.
\\ \ \\
\noindent {\bf LEP bounds.}
Dark matter with $ m_\chi > m_t $ cannot be produced on--shell at LEP.
However, there are still significant constraints on Z$^\prime$ due
to the effective operators
\begin{equation}
{g_f^2 \over m_{Z'}^2}~ \bar f_i \gamma^\mu \gamma^5 f_i ~~  \bar f_j \gamma_\mu \gamma^5 f_j ~,
\end{equation}
for various fermions $f_i$ and $f_j$. These operators are constrained 
by  the precise measurements of the cross sections and angular distributions of the  final state fermions. In the axial case, the resulting bound is 
\cite{Alcaraz:2006mx}
\begin{equation}
 {m_{Z'} \over g_f  } > 5 ~{\rm TeV} ~.
 \end{equation}
Comparing this to Eq.~(\ref{annih}), one finds that Z$^\prime$ couples
much stronger to DM than it does to SM fermions, 
\begin{equation}
\alpha = g_\chi/ g_f > 10 ~.
 \end{equation}
\\ \ \\
\noindent {\bf Perturbativity.}
As is clear from the above discussion, the DM--Z$^\prime$ coupling can become quite
strong. Then, our approximation  is controllable only if
 \begin{equation}
{g_\chi^2 \over 4 \pi^2} < 1   ~.
\label{pert}
 \end{equation}
We do not impose further constraints on the position of the Landau pole 
of the coupling as we expect our model to be UV--completed already in 
the multi--TeV range.
\\ \ \\
\noindent {\bf Dilepton and monojet LHC bounds.}
At the LHC, both dark matter and Z$^\prime$  can be produced on--shell,
which leads to strong bounds from CMS and ATLAS.
The most important constraint is due to searches for dileptons with a large 
invariant mass. 
We will use the CMS Z$^\prime$ analysis of 3.6 fb$^{-1}$/8 TeV and   
5 fb$^{-1}$/7 TeV  \cite{Chatrchyan:2012oaa}  as our benchmark constraint. The result is summarized
in Fig.2 (upper right panel) of that paper. For a sequential SM  Z$^\prime$
(SSM), that is having the same couplings as the Standard Model  Z-boson, the 
exclusion limit is around 2.5 TeV. To adapt the results to our case,
one must take into account the difference in the Z$^\prime$ couplings as well as the reduced branching ratio for  Z$^\prime$ decay into visible fermions,
\begin{equation}
\sigma_{l^{+}l^{-}} \rightarrow   \left(  {g_f \over g_Z}  \right)^2   {\rm BR_{\rm vis}} ~\sigma_{l^{+}l^{-} }
\end{equation}
For our estimates it suffices to approximate $g_Z$ by its 
(universal) axial component, $g_2/(4 \cos \theta_W)$. The 
branching ratio for  Z$^\prime$ decay into SM fermions is 
\begin{equation}
{\rm BR_{\rm vis}} \simeq { 45 g_f^2 \over 45 g_f^2 + g_{\chi }^2 \beta^3} ~,
\end{equation}
where  $\beta = \sqrt{1- 4 m_\chi^2 / m_{Z'}^2}$ accounts
for the kinematic suppression  in
an axial--vector decay (see e.g. \cite{Heinemeyer:2007bw}).
These factors result in the dependence of the number of expected 
$l^{+}l^{-}$ events on $g_f$ and $ g_{\chi }$. The constraints on 
Z$^\prime$ relax significantly as $g_f$ decreases and $m_{Z'}$
as light as 500 GeV becomes allowed given it decays predominantly invisibly.

 To estimate the 
resulting LHC bound on $m_{Z'}$, we analytically approximate  the  $l^{+}l^{-}$ production  cross section in Fig.2 of \cite{Chatrchyan:2012oaa} and calculate how much it should be reduced
to comply with its experimental bound. We find that the result
can be cast in the form
$ m_{Z'} > m_0 + 0.55 \log_{10} \left[ (g_f/0.17)^2 ~{\rm BR_{\rm vis}}
   \right] $, 
 for $m_{Z'} $ in TeV  and $m_0$ being an $m_{Z'}$--range dependent 
 constant:  $m_0 \simeq (2,2.3,2.5)$ TeV for $m_{Z'}\sim (0.5,1, \geq 1.5) $ TeV.
 For instance,   a 500 GeV Z$^\prime$ becomes  allowed if the $l^{+}l^{-}$
 cross section reduces by about 3 orders of magnitude, whereas a 2.5 TeV
 Z$^\prime$ is allowed with no suppression required.
 
Z$^\prime$ models are also constrained by monojet events with large missing energy, which is due to Z$^\prime$ decay into dark matter. 
The ATLAS analysis of 10.5 fb$^{-1}$/8 TeV data \cite{ATLAS:2012zim}
 imposes the bound
on the axial--vector interaction (D8--operator of \cite{Goodman:2010ku}),
\begin{equation}
{m_{Z'} \over \sqrt{ g_f g_\chi} } > 600-700 ~{\rm GeV} ~,
\end{equation}
for $m_\chi \sim 200$ GeV (and a weaker bound for heavier DM).
Inclusion of on-shell effects does not make the constraint significantly
stronger \cite{Arcadi:2013qia}. As a result,
similarly to the direct DM detection constraint, it is satisfied 
when Eq.~(\ref{annih}) is imposed. 
\\ \ \\
\noindent {\bf Combined constraints.}
The above estimates serve to single out the most important constraints,
whose compatibility is to be analyzed. We see that,
once the correct DM relic abundance is imposed, the main factors restricting
available parameter space are the LHC dilepton bound and perturbativity.
Indeed, the LHC constraint can always be satisfied by decreasing $g_f$,
which according to Eq.~(\ref{annih}) increases $g_\chi$ until it
hits the perturbative bound (\ref{pert}). We find that all of the constraints are in fact compatible.  For instance, at  $m_{Z'} \sim
500$ GeV, the allowed range of $\alpha= g_\chi/ g_f$ 
spans about two orders of magnitude, from 
${\cal O}(10)$ to ${\cal O}(10^3)$.

\begin{figure}
  \centering
  \subfigure[]{\includegraphics[height=20em]{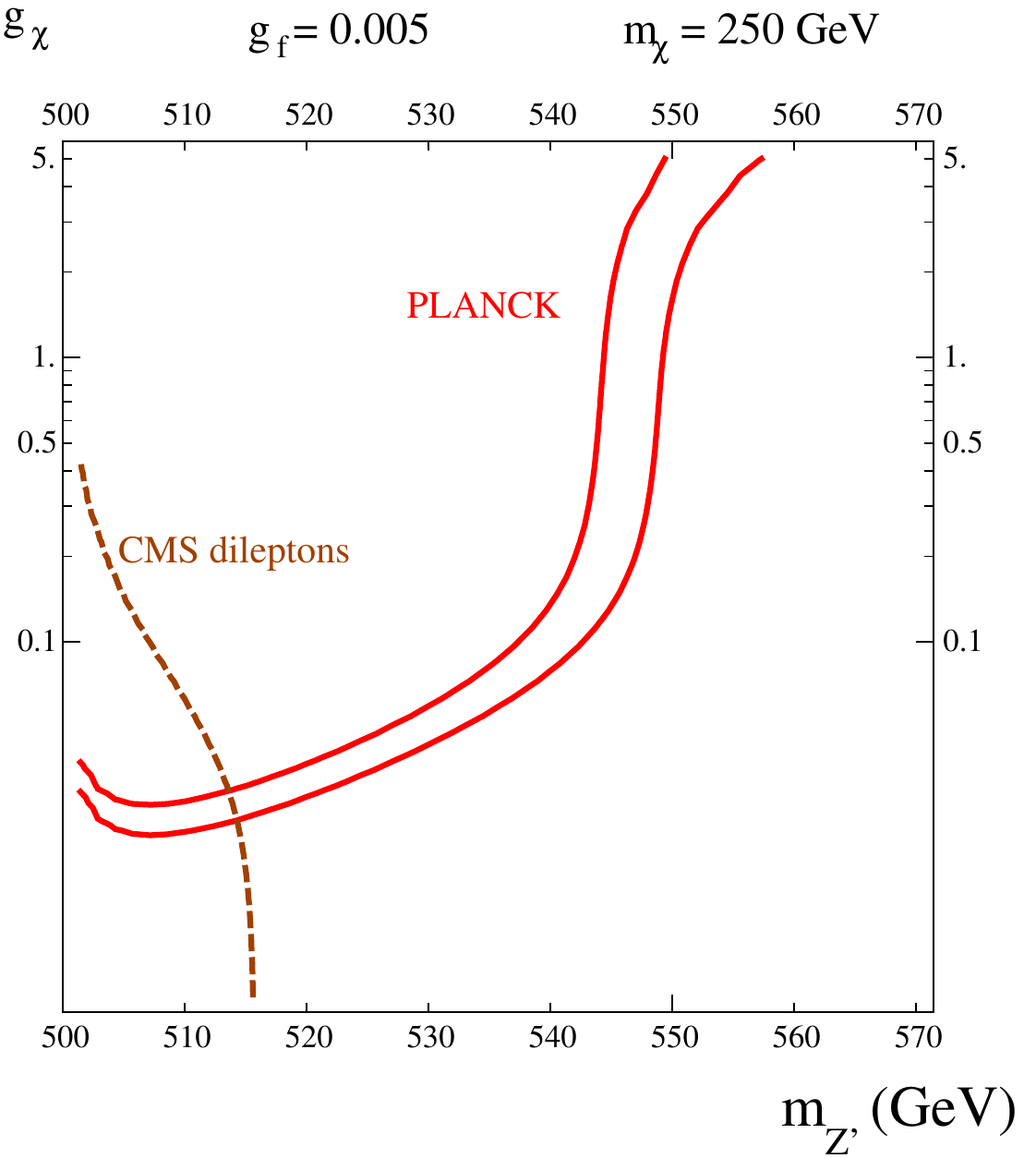}}
  \subfigure[]{\includegraphics[height=20em]{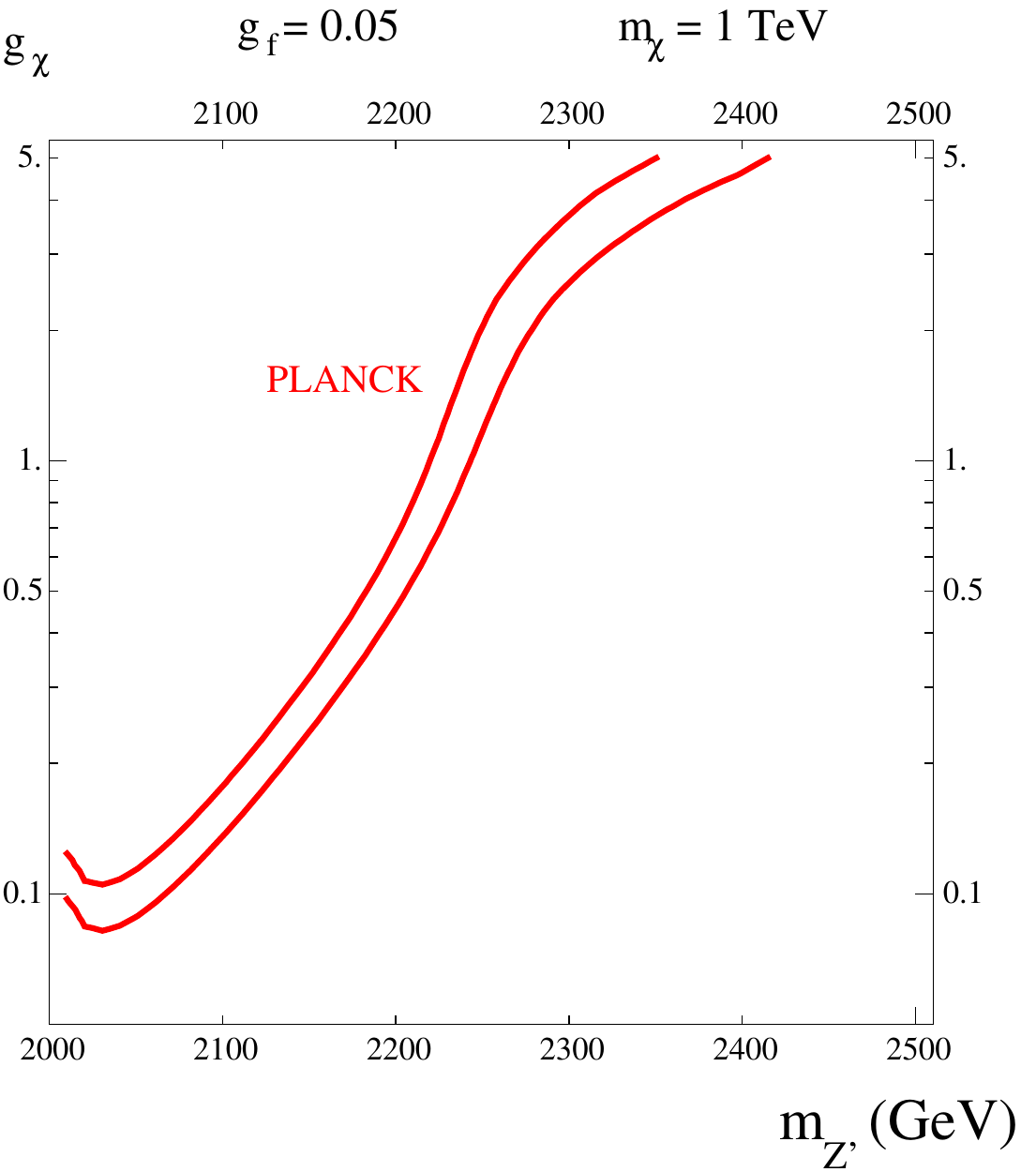}}
  \caption{Relic density (Planck) and direct search (CMS dileptons) constraints on Z$^\prime$ for lighter (left panel) and
  heavier (right panel) dark matter. The area between the two red (solid) lines is consistent with Planck, while the area below the brown (dashed) line is excluded by CMS.  Other constraints (LEP, direct DM detection, monojets) are satisfied automatically.}
  \label{fig}
\end{figure}

To go further, let us remind the reader that
our estimate  of the DM relic abundance constraint (\ref{annih}) 
is rather simplistic. It does not take into account resonant effects
nor  those due to  thermal averaging.
The correct treatment is  provided by the numerical package micrOMEGAs
  \cite{Belanger:2013oya}.
 Using this tool, we find the same qualitative conclusion:
 all of the constraints are compatible.  
In our numerical study, we impose the condition\footnote{This condition may not be necessary if the coupling $g_f$ is very small and
the Z$^\prime$ production cross section is suppressed altogether.}
\begin{equation}
m_{Z'} > 2 m_\chi ~,
\end{equation} 
which allows for the Z$^\prime$ decay into dark matter and 
amounts to ``invisibility'' of the former.
    Two representative results are shown in Fig.~1. In the left
   panel, we set  $m_\chi= 250$ GeV, $g_f=0.005$ and scan parameter space
   $\{ m_{Z'}, g_\chi \}$  satisfying the PLANCK constraint \cite{Ade:2013ktc}.  
 As $m_{Z'}$ increases from $2 m_\chi $, so does $g_\chi$. In this region,
 the effect of resonant annihilation is important. The resonance is quite
 broad, on the order of tens of GeV, due to the thermal smearing. 
 Away from the resonance,  $g_\chi$ quickly turns non--perturbative. 
  The LHC constraint excludes part of the parameter space close to the threshold,
 where the invisible Z$^\prime$ decay is inefficient.
Further constraints from LEP, direct DM detection and  monojets
are satisfied automatically in this panel.
The result is that the $m_{Z'}$  range 520-560 GeV is allowed, while 
$g_\chi$ varies by two orders of magnitude,  from 0.05 to 5.

For heavier DM,  the LHC bound becomes less severe and
the coupling $g_f$ is allowed to be larger. 
For example, in the right panel of Fig.~1, the CMS constraint is satisfied everywhere due to  the suppressed Z$^\prime$
production with $g_f=0.05$. The  $m_{Z'}$ range satisfying 
PLANCK extends over hundreds of GeV.

The pattern observed in this figure is quite general: the allowed parameter space 
is not far  from the resonance region, $m_{Z'} \gsim  2 m_\chi$,
with the latter being relatively broad, $\sim 10-20 \% ~m_{Z'} $. The LEP, direct DM detection and  monojet constraints
are satisfied automatically in the region of interest,
while the dilepton LHC bound cuts out part of the parameter space.
The dark matter candidate, AxDM, belongs to the general WIMP category as it has a TeV scale mass and couplings  in
the range ${\cal O}(10^{-2}-1)$.
 A detailed scan of AxDM parameter space  is reserved for 
 a subsequent publication.

\section{Conclusion}
In this paper, we have explored a very simple scenario in which
dark matter couples to the SM fermions via an axial Z$^\prime$
(``axial dark matter''). The model is characterized by 2 couplings as
well as  the Z$^\prime$ and DM masses. If the  Z$^\prime$ couples much stronger to the dark sector
compared to the visible sector, which may be due to a mixing of two U(1)'s, all the phenomenological constraints  
can be satisfied. In particular, the LHC constraints are loose due to invisible Z$^\prime$ decay and allow for $m_{Z'}$ as low as 500 GeV, while DM scattering on nuclei is spin--dependent
and thus weakly constrained. The correct DM relic density is obtained
in regions not far from the resonance,   $m_{Z'} \gsim  2 m_\chi$.
All in all, we find that AxDM is consistent with the  thermal WIMP paradigm.  

{\bf Acknowledgements.} The authors would like  to thank G. Arcadi, B. Zaldivar, M. Tytgat and K. Tarachenko  for very useful discussions.
This  work was supported by the French ANR TAPDMS {\bf ANR-09-JCJC-0146}
and the Spanish MICINN's Consolider-Ingenio 2010 Programme  under grant
Multi-Dark {\bf CSD2009-00064}.
Y.M.  acknowledges partial support from the European Union FP7 ITN
INVISIBLES (Marie Curie Actions, PITN- GA-2011- 289442) and from the ERC
advanced grant Higgs@LHC.


\begin{thebibliography}{99}

\bibitem{Langacker:2008yv}
  P.~Langacker,
  Rev.\ Mod.\ Phys.\  {\bf 81} (2008) 1199;
  T.~Han, P.~Langacker, Z.~Liu and L.~-T.~Wang,
  arXiv:1308.2738 [hep-ph].



\bibitem{Domingo:2013tna} 
  F.~Domingo, O.~Lebedev, Y.~Mambrini, Jérém.~Quevillon and A.~Ringwald,
  JHEP {\bf 1309}, 020 (2013).

 
\bibitem{Cheung:2007ut} 
  K.~Cheung and T.~-C.~Yuan,
  JHEP {\bf 0703}, 120 (2007);
Y.~Mambrini,
  JCAP {\bf 1107} (2011) 009;
  V.~Barger, D.~Marfatia and A.~Peterson,
  Phys.\ Rev.\ D {\bf 87}, 015026 (2013);
 E.~Dudas, L.~Heurtier, Y.~Mambrini and B.~Zaldivar,
  JHEP {\bf 1311}, 083 (2013).


\bibitem{Bertone:2004pz} 
  G.~Bertone, D.~Hooper and J.~Silk,
  Phys.\ Rept.\  {\bf 405}, 279 (2005).


\bibitem{Chatrchyan:2012oaa} 
  S.~Chatrchyan {\it et al.}  [CMS Collaboration],
  Phys.\ Lett.\ B {\bf 720}, 63 (2013).
   
\bibitem{Aad:2012hf} 
  G.~Aad {\it et al.}  [ATLAS Collaboration],
  JHEP {\bf 1211}, 138 (2012).

\bibitem{Akerib:2013tjd} 
  D.~S.~Akerib {\it et al.}  [LUX Collaboration],
  arXiv:1310.8214 [astro-ph.CO].


\bibitem{Arcadi:2013qia} 
  G.~Arcadi, Y.~Mambrini, M.~H.~G.~Tytgat and B.~Zaldivar,
  arXiv:1401.0221 [hep-ph].

\bibitem{Alves:2013tqa} 
  A.~Alves, S.~Profumo and F.~S.~Queiroz,
  arXiv:1312.5281 [hep-ph].


\bibitem{deSimone:2014pda}
  A.~De Simone, G.~F.~Giudice and A.~Strumia,
  arXiv:1402.6287 [hep-ph].



\bibitem{Holdom:1985ag} 
  B.~Holdom,
  Phys.\ Lett.\ B {\bf 166}, 196 (1986).
 
\bibitem{Aprile:2013doa} 
  E.~Aprile {\it et al.}  [XENON100 Collaboration],
  Phys.\ Rev.\ Lett.\  {\bf 111}, no. 2, 021301 (2013);
S.~Archambault {\it et al.}  [PICASSO Collaboration],
  Phys.\ Lett.\ B {\bf 711}, 153 (2012);
E.~Behnke {\it et al.}  [COUPP Collaboration],
  Phys.\ Rev.\ D {\bf 86}, 052001 (2012).




\bibitem{Freytsis:2010ne} 
  M.~Freytsis and Z.~Ligeti,
  Phys.\ Rev.\ D {\bf 83}, 115009 (2011).

\bibitem{Boehm:2014hva}
  C.~Boehm, M.~J.~Dolan, C.~McCabe, M.~Spannowsky and C.~J.~Wallace,
  arXiv:1401.6458 [hep-ph].


\bibitem{Alcaraz:2006mx} 
  J.~Alcaraz {\it et al.}  [ALEPH and DELPHI and L3 and OPAL and LEP Electroweak Working Group Collaborations],
  hep-ex/0612034.






\bibitem{Heinemeyer:2007bw} 
  S.~Heinemeyer, W.~Hollik, A.~M.~Weber and G.~Weiglein,
  JHEP {\bf 0804}, 039 (2008).
 

\bibitem{ATLAS:2012zim} 
  [ATLAS Collaboration],
  ``Search for New Phenomena in Monojet plus Missing Transverse Momentum Final States using 10fb-1 of pp Collisions at sqrt{s}=8 TeV with the ATLAS detector at the LHC,''
  ATLAS-CONF-2012-147.


\bibitem{Goodman:2010ku} 
  J.~Goodman, M.~Ibe, A.~Rajaraman, W.~Shepherd, T.~M.~P.~Tait and H.~-B.~Yu,
  Phys.\ Rev.\ D {\bf 82}, 116010 (2010).



\bibitem{Belanger:2013oya} 
  G.~Belanger, F.~Boudjema, A.~Pukhov and A.~Semenov,
  Comput.\ Phys.\ Commun.\  {\bf 185}, 960 (2014).



\bibitem{Ade:2013ktc} 
  P.~A.~R.~Ade {\it et al.}  [Planck Collaboration],
  arXiv:1303.5062 [astro-ph.CO].




\end{thebibliography}
\end{document}